# Dirac-Point Solitons in Nonlinear Optical Lattices

Kang Xie[1,*], Qian Li[1], Allan D. Boardman[2], Qi Guo[3], Zhiwei Shi[4], Haiming Jiang[1,*], Zhijia Hu[1*], Wei Zhang[1], Qiuping Mao[1], Lei Hu[1], Tianyu Yang[1], Fei Wen[1], Erlei Wang[1]

[1]School of Instrument Science and Opto-electronic Engineering, Hefei University of Technology, Hefei, 230009, P.R. China

[2]Joule Physics Laboratory, Institute for Materials Research, University of Salford, Salford, Manchester M5 4WT, UK

[3]Guangdong Provincial Key Laboratory of Nanophotonic Functional Materials and Devices, School of Information and Optoelectronic Science and Engineering, South China Normal University, Guangzhou, 510631, P.R. China

[4]Faculty of Information Engineering, Guangdong University of Technology, Guangzhou 510006, P.R. China

**Abstract:**

The discovery of a new type of solitons occuring in periodic systems without photonic bandgaps is reported. Solitons are nonlinear self-trapped wave packets. They have been extensively studied in many branches of physics. Solitons in periodic systems, which have become the mainstream of soliton research in the past decade, are localized states supported by photonic bandgaps. In this Letter, we report the discovery of a new type of solitons located at the Dirac point beyond photonic bandgaps. The Dirac point is a conical singularity of a photonic band structure where wave motion obeys the famous Dirac equation. These new solitons are sustained by the Dirac point rather than photonic bandgaps, thus provides a sort of advance in conceptual understanding over the traditional gap solitons. Apart from their theoretical impact within soliton theory, they have many potential uses because such solitons have dramatic stability characteristics and are possible in both Kerr material and photorefractive crystals that possess self-focusing and self-defocusing nonlinearities. The new results elegantly reveal that traditional photonic bandgaps are not required when Dirac points are accessible. The findings enrich the soliton family and provide valuable information for studies of nonlinear waves in many branches of physics, including hydrodynamics, plasma physics, and Bose–Einstein condensates.

PACS numbers:42.65.Tg, 42.70.Qs, 42.82.Et, 73.22.Pr, 78.67.Pt

Confinement of waves within a finite area is the basis of information processing [1]. Traditionally, wave trapping is achieved by cavities and waveguides that rely on total internal reflection, or photonic bandgap, to suppress radiation losses [2, 3]. Cavities and waveguides can be formed by a high-index core surrounded by a cladding with a lower refractive index so that total internal reflection can take place [2]. Alternatively, cavities and waveguides can be formed in periodic systems as defects [3]. Due to the presence of allowed bands and forbidden gaps, radiation losses of the wave accommodated by the defect are suppressed by any photonic bandgap of the systems. There are various other ways and, among them, nonlinearity is an unique example. Localized modes due to nonlinearity are commonly called solitons [4, 5]. In homogeneous media, nonlinearity raises refractive index of the media so that light creates its own high-index core. In this way light is essentially guided by total internal reflection. In periodic systems, nonlinearity changes its onsite refractive index so that light creates its own defect. In this case light is trapped by photonic bandgaps of the periodic lattices. Solitons are nonlinear self-trapped wave packets. They have been extensively studied in many branches of physics. Solitons in periodic systems have become the mainstream of soliton research in the past decade [6, 7].

During the same period, research on graphene has made great progress [8]. The electronic band structure of graphene contains Dirac cones at the six corners of the hexagonal Brillouin zone. The associated energy-wavenumber relation resembles the two-dimensional massless Dirac equation $-iv(\sigma_x \partial_x + \sigma_y \partial_y)\Psi = (\omega - \omega_D)\Psi$ for relativistic electrons in a vacuum, where $v$ is the velocity, $\omega_D/2\pi$ is the Dirac frequency, $\sigma_x$ and $\sigma_y$ are Pauli matrices, and $|\Psi|^2$ is the probability of finding the spinors in space. Building on the observations of graphene, it is found that the band structure of a photonic crystal formed by a two-dimensional triangular lattice also possesses Dirac cones at the corners of the Brillouin zones [9]. At these high-symmetry points Maxwell's equations can be replaced by the massless Dirac equation with $\Psi$ being the wave functions of two degenerate Bloch states. Wave behaviour at the Dirac frequency has been studied extensively in photonic crystals since then [10-12], and localized modes have been found recently at the Dirac frequency [13, 14]. It has been shown that the Dirac point in band structures of these lattices can take the role of a bandgap to form localized modes at a defect, a mechanism different from that of a nonlinear Dirac soliton of the nonlinear relativistic Dirac equation [15]. Soliton in photonic crystals is essentially a nonlinearity-induced defect mode, so it is natural to ask if such a self-localized mode can be supported, or not, by the same Dirac point.

In this Letter, we report the discovery of a new type of solitons occuring at the Dirac point. It is found that, besides photonic bandgaps, a Dirac point in the band structure of a triangular nonlinear lattice can also sustain self-localized nonlinear modes. This new specific entity is designated here as Dirac-point soliton. We show that such solitons are possible in both Kerr material and photorefractive crystals with self-focusing and self-defocusing nonlinearities. Characteristics of the Dirac-point solitons are revealed and their stability condition is analyzed by linear stability analysis. It is found that the Dirac-point solitons satisfy the so-called Vakhitov-Kolokolov stability criterion [16]. We verify the stability criterion by direct numerical simulations.

The propagation of optical beams in a nonlinear periodic array is described by the nonlinear Schrödinger equation for the slowly varying amplitude of the light [17, 18]:

$$i\frac{\partial U}{\partial Z} + \left(\frac{\partial^2}{\partial X^2} + \frac{\partial^2}{\partial Y^2}\right)U - V_{NL}U = 0 \qquad (1)$$

where the wave is presumed to propagate predominantly along the Z-direction. The potential for a Kerr nonlinearity is $V_{NL} = V - \sigma|U|^2$, where $\sigma=1$ (or $-1$) corresponds to a Kerr self-focusing (or self-defocusing) nonlinearity. The linear index potential $V = V_0[\chi + \cos(b_1 \cdot r) + \cos(b_2 \cdot r) + \cos(b_3 \cdot r)]^2$ represents a triangular lattice, where $b_1 = 2\pi\left(\hat{x} - \frac{\sqrt{3}}{3}\hat{y}\right)$ and $b_2 = 2\pi\left(0\hat{x} + \frac{2\sqrt{3}}{3}\hat{y}\right)$ are the reciprocal lattice basis vectors, and $b_3 = 2\pi\left(-\hat{x} - \frac{\sqrt{3}}{3}\hat{y}\right)$. The potential for a saturable nonlinearity is $V_{NL} = V_0/(1+I+|U|^2)$, which suits the description of photorefractive crystals. If $V_0>0$ (<0) the medium nonlinearity has a self-focusing (self-defocusing) nature. The normalized intensity pattern $I(X,Y) = I_0[\chi + \cos(b_1 \cdot r) + \cos(b_2 \cdot r) + \cos(b_3 \cdot r)]^2$ can be generated experimentally on a stationary background by interfering three plane waves with intensity $I_0$ and transverse wave vectors $b_i$. The three wave vectors of the plane waves form a triangle, i.e., $b_1+b_2+b_3=0$.

In the linear limit Eq. (1) reduces to $i\frac{\partial U}{\partial Z} + \left(\frac{\partial^2}{\partial X^2} + \frac{\partial^2}{\partial Y^2}\right)U - VU = 0$. Wave propagation in such a linear periodic lattice has the form of $U = \Phi(X,Y)e^{-iqZ}$ and is known to exhibit unique features that arise from the presence of allowed bands and forbidden gaps. The band structure of the lattice, which can be found by the plane wave expansion method [1], exhibits Dirac cones at the six corners of the Brillouin zone at an eigenvalue $q_D$. At the Dirac point $q=q_D$, the density of radiation states is precisely zero [20], which means that outgoing waves are forbidden in the surrounding medium. Because of this feature, field concentration around a defect becomes possible and the optical lattice can support localized modes at the Dirac point. Examples of the potential $V$ and its associated linear defect-guided modes are discussed in Supplemental Material. There exists a range of $q$, around $q_D$, where decay rate is low and optical wave-guiding by a defect is practically realizable. The defect could be created by nonlinearly-induced onsite index change. If the defects are self-induced optically by nonlinearity, the corresponding, self-localized nonlinear modes are referred to as solitons. In other words, the presence of a defect mode at the Dirac point in the linear limit suggests the existence of a Dirac-point soliton in the corresponding nonlinear system. This is, indeed, the case. To find the Dirac-point solitons of Eq. (1), we seek a solution of the form $U(X,Y,Z) = \Phi(X,Y)e^{-iqZ}$. Given this substitution, Eq. (1) reduces to

$$\left(\frac{\partial^2}{\partial X^2} + \frac{\partial^2}{\partial Y^2}\right)\Phi - V_{NL}\Phi = -q\Phi \qquad (2)$$

Eq. (2), which is a static nonlinear Schrödinger equation, is solved numerically by the modified squared-operator method [21] for solitary wave solutions. For a fundamental soliton $\Phi$ is real, while for a vortex soliton $\Phi = f(r)e^{im\theta}$, where $m$ is an integer. Typical Dirac-point solitons and the power $P = \iint |\Phi|^2 \, dXdY$ conveyed by the solitons are shown in Figs. 1 and 2 for the Kerr self-focusing and self-defocusing nonlinearities. Results for Dirac-point solitons in photorefractive crystals are presented in Figs. S5 and S6 [19]. The gray scale in the background of Figs. 1(c, d) and 2(c, d) indicates the level of linear losses $\alpha$ of the wave [19] in the lattice potential. The Dirac-point solitons are found to exist within the ranges shaded light gray, where the level of the linear decay rate $\alpha$ is low.

The Dirac-point solitons in a nonlinear media preserve their shape, but their stability is not guaranteed, because of the non-integrable nature of the underlying equation. In fact, their stability is a crucial issue because only stable (or weakly unstable) modes can be observed experimentally. To study the stability of these solitons, a perturbation of the form $U = e^{-iqZ}\left[\Phi + (v-w)e^{\lambda Z} + (v+w)^* e^{\lambda^* Z}\right]$, is invoked, where $v, w \ll 1$. Substituting this perturbation into Eq. (1) and then linearizing, results in the need to solve the linear stability eigenvalue problem $L\Psi = \lambda\Psi$, with $\Psi$ being the transpose of $(v, w)$ and

$$L = -i\begin{pmatrix} \frac{1}{2}\sigma(\Phi^{*2} - \Phi^2) & \Delta + q - V + 2\sigma|\Phi|^2 - \frac{1}{2}\sigma(\Phi^2 + \Phi^{*2}) \\ \Delta + q - V + 2\sigma|\Phi|^2 + \frac{1}{2}\sigma(\Phi^2 + \Phi^{*2}) & \frac{1}{2}\sigma(\Phi^2 - \Phi^{*2}) \end{pmatrix}$$

for Kerr nonlinearity [19]. The operator is $\Delta = \frac{\partial^2}{\partial X^2} + \frac{\partial^2}{\partial Y^2}$. The eigenvalue problem

$L\Psi=\lambda\Psi$ is solved by a numerical iteration method [22]. The real part of the perturbation growth rate Re($\lambda$) versus the propagation constant $q$ of the soliton is plotted in Figs. 1(d) and 2(d) for the Kerr lattice and Figs. S5(d) and S6(d) for the photorefractive lattice [19]. In a saturable self-focusing lattice (Fig. S5), growth rates exceed 10, both the fundamental and the first vortex solitons are unstable. In a saturable self-defocusing lattice (Fig. S6), growth rates are of the order of 1, both the fundamental and the first vortex solitons are weakly unstable. In Kerr lattices (Figs. 1 and 2), the fundamental soliton is unstable for self-focusing nonlinearity and stable for self-defocusing nonlinearity. The power curves in Figs. 1(c) and 2(c) also give information on the stability of the solitons. According to the Vakhitov-Kolokolov stability criterion [16], a soliton is stable (unstable) if slope of the corresponding power curve is positive (negative). The Vakhitov-Kolokolov stability criterion is derived for homogenous nonlinear medium but there is a periodic potential present for the Dirac-point solitons, so it is not directly applicable. However, our linear stability analysis comes up with results that agree well with the Vakhitov-Kolokolov stability criterion, despite the periodicity of the potential. More specifically, the Dirac-point solitons in self-defocusing lattices [Figs. 1(b) and 2(b), Figs. S6(a) and S6(b)] exhibit $dP/dq \geq 0$ on the power curves and are stable (or weakly unstable), whereas the Dirac-point solitons in self-focusing lattices [Figs. 1(a) and 2(a), Figs. S5(a) and S5(b)] exhibit $dP/dq < 0$ and are unstable.

The stability of Dirac-point solitons can be checked numerically using the split-step Fourier method. Simulated evolution scenarios of typical stable and unstable solitons are shown in Figs. 3 and 4 respectively, confirming the stability analysis. The propagation distance of the stable soliton (Fig. 4) is about three orders of magnitude larger than that of the unstable soliton (Fig. 3). The Dirac-point solitons belong to the group of algebraic solitons [23]. As shown in Figs. 3(e) and 4(e), tail of the Dirac-point soliton decays algebraically according to a power-law (roughly $r^{-3/2}$) at large distances. This is understandable because the field is so weak in the cladding that it resumes a linear behaviour, and in the linear limit a triangular optical lattice can support waves with power-law asymptotics at the Dirac point [13, 14]. The spectrum of the stable soliton is shown in Fig. 4(f), which verifies that the propagation constant of the Dirac-point soliton is truly centered at the Dirac point $q_D=-41.81$. On the other hand, the propagation constant of the unstable soliton [Fig. 3(f)] sweeps a wide range of values. As the Dirac-point soliton breaks down, its amplitude reduces. This alters parameters of the nonlinearity-induced defect, and, in turn, shifts its propagation constant.

In summary, a new type of solitons, which rely on the Dirac point rather than photonic bandgaps to establish field localization, are discovered in photonic lattices. The Dirac equation is a special symbol of relativistic quantum mechanics, from merging quantum mechanics with special relativity to predicting the existence of anti-matter. Investigations in the Dirac-point solitons may lead to new findings in many relativistic quantum effects on the transport of photons, phonons, and electrons.

This work is supported by the National Science Foundation of China (11574070)

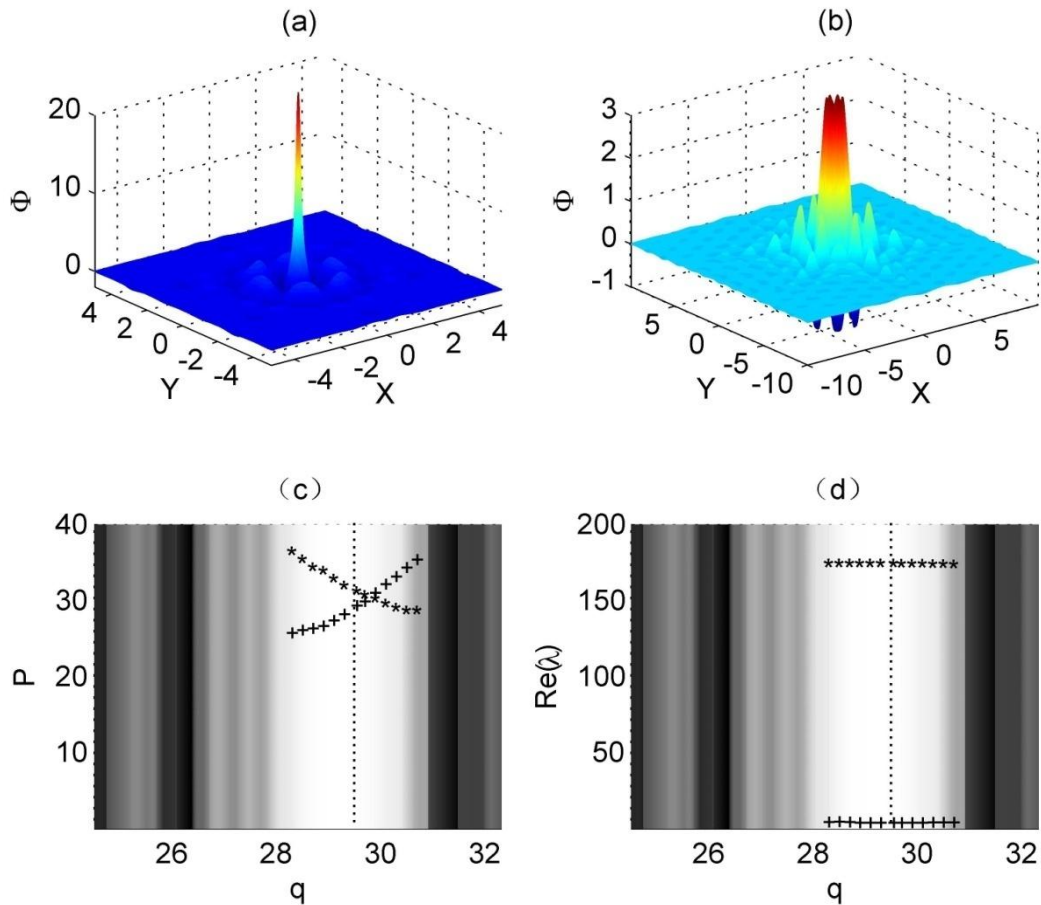

FIG. 1. Dirac-point solitons in Kerr nonlinear media. The lattice potential is $V = 10[3/2 + \cos(b_1 \cdot r) + \cos(b_2 \cdot r) + \cos(b_3 \cdot r)]^2$, which exhibits absolute index maxima on lattice sites. (a, b) The field profiles of the solitons in a Kerr self-focusing (a) and self-defocusing (b) lattice at $q_D=29.445$. (c, d) Power $P$ (c) and real part of $\lambda$ (d) versus $q$, where "*" corresponds to the self-focusing case and "+" corresponds to the self-defocusing case. The dotted vertical line indicates the position of the Dirac point.

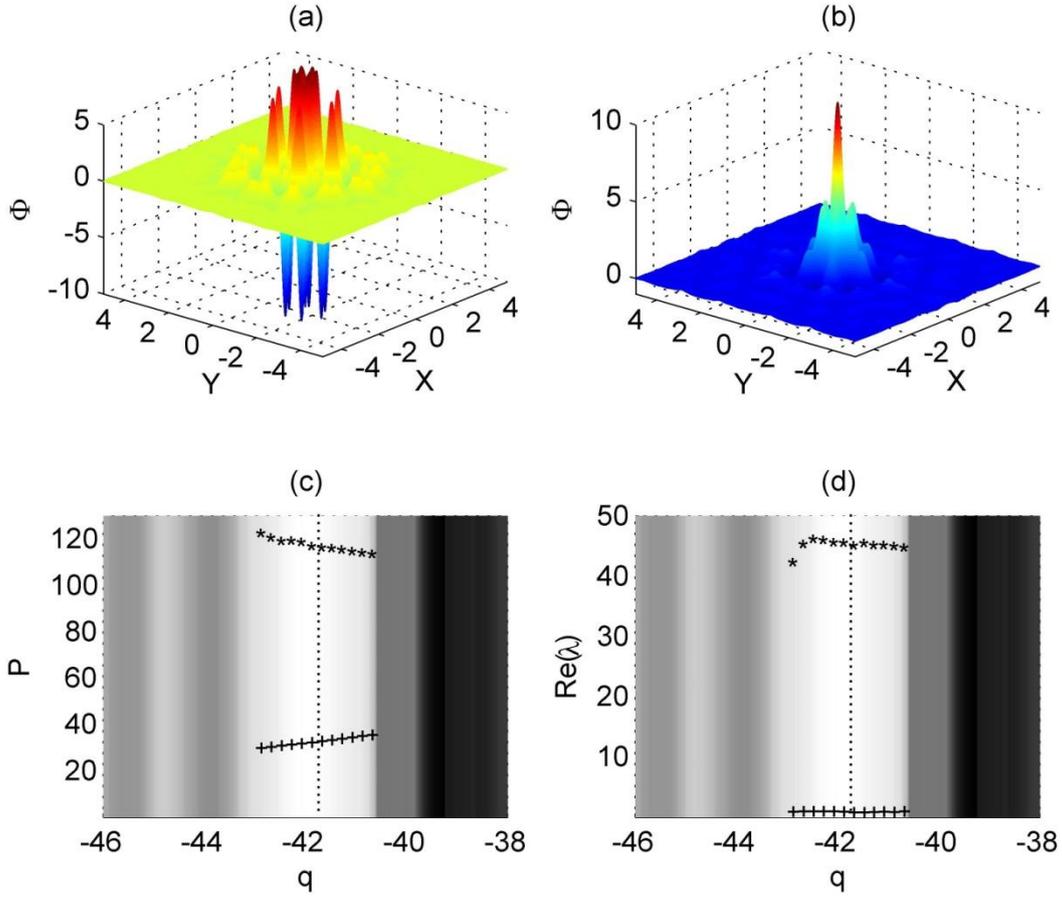

FIG. 2. Dirac-point solitons in Kerr nonlinear media. The lattice potential is
$V = -35[-1/3 + \cos(b_1 \cdot r) + \cos(b_2 \cdot r) + \cos(b_3 \cdot r)]^2$, which exhibits absolute index minima on lattice sites. (a, b) The field profiles of the solitons in a Kerr self-focusing (a) and self-defocusing (b) lattice at $q_D = -41.81$. (c, d) Power $P$ (c) and real part of $\lambda$ (d) versus $q$, where "*" corresponds to the self-focusing case and "+" corresponds to the self-defocusing case. The dotted vertical line indicates the position of the Dirac point.

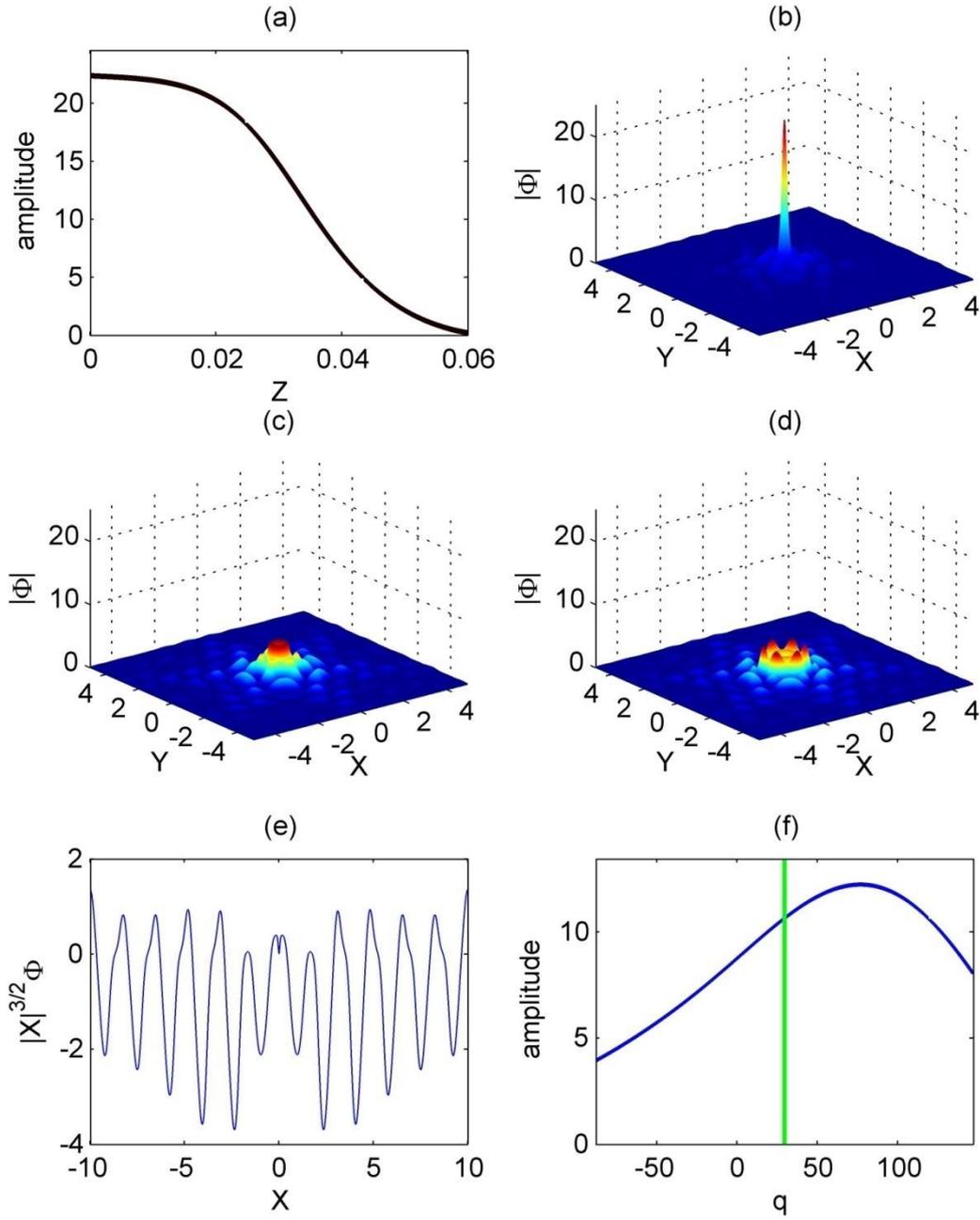

FIG. 3. Breakdown of the Dirac-point soliton in a Kerr self-focusing lattice. The lattice potential is $V = 10[3/2 + \cos(b_1 \cdot r) + \cos(b_2 \cdot r) + \cos(b_3 \cdot r)]^2$ and the initial profile of the soliton is shown in Fig. 1(a). (a) Evolution of the soliton amplitude $|\Phi(0,0)|$. (b-d) The $|\Phi|$ field at respectively $Z=0.006, 0.048, 0.06$. (e) Product of the initial $\Phi$ and $r^{3/2}$ on the $X$ axis. (f) Propagation constant spectrum of the soliton. The green vertical line indicates the position of the Dirac point.

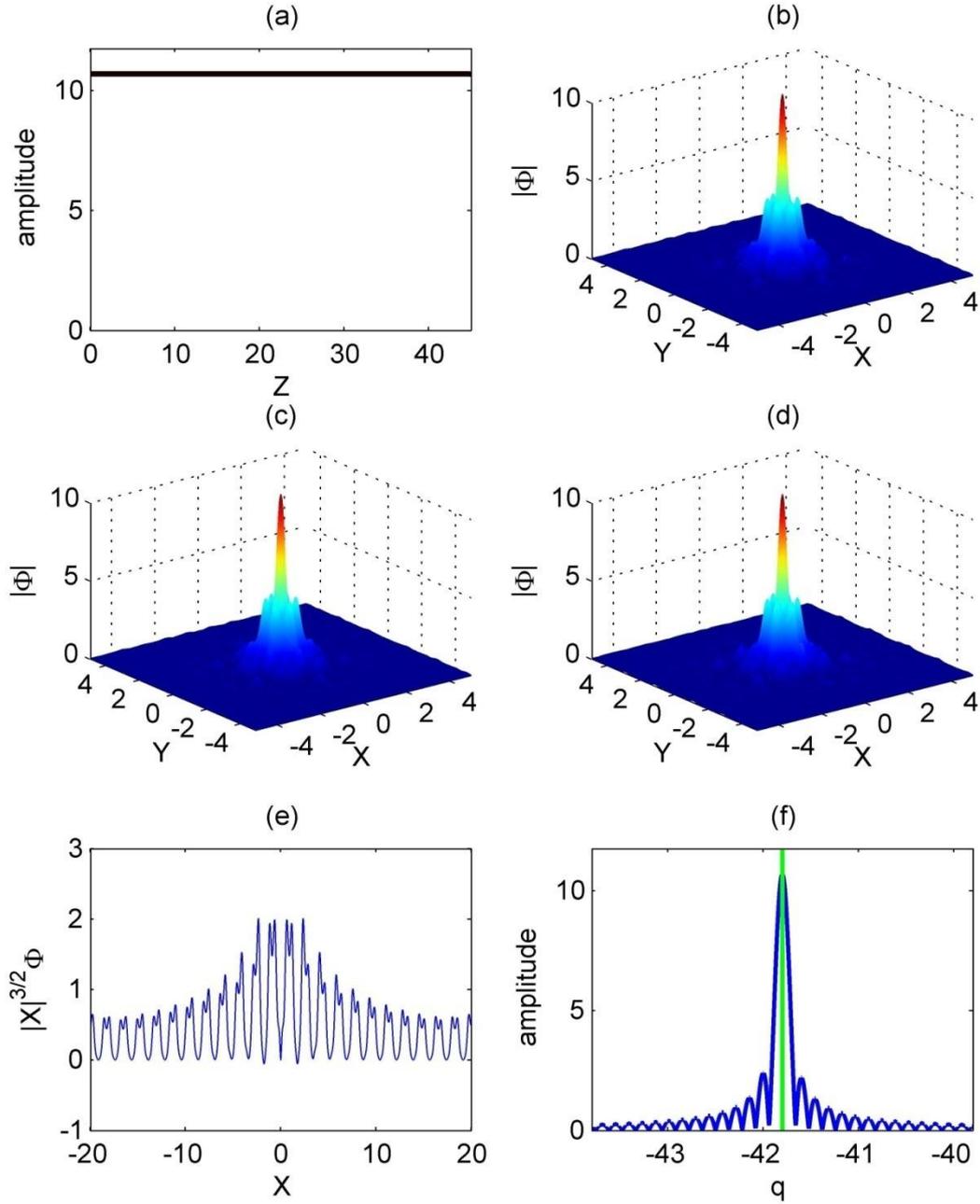

FIG. 4. Dynamics of the Dirac-point soliton in a Kerr self-defocusing lattice. The lattice potential is $V = -35[-1/3 + \cos(b_1 \cdot r) + \cos(b_2 \cdot r) + \cos(b_3 \cdot r)]^2$ and the initial profile of the soliton is shown in Fig. 2(b). (a) Evolution of the soliton amplitude $|\Phi(0,0)|$. (b-d) The $|\Phi|$ field at respectively $Z=4.5, 22.5, 45$. (e) Product of the initial $\Phi$ and $r^{3/2}$ on the $X$ axis. (f) Propagation constant spectrum of the soliton. The green vertical line indicates the position of the Dirac point.

# Supplemental Material for
## " Dirac-Point Solitons in Nonlinear Optical Lattices "

Kang Xie, Qian Li, Allan D. Boardman, Qi Guo, Zhiwei Shi, Haiming Jiang, Zhijia Hu, Wei Zhang, Qiuping Mao, Lei Hu, Tianyu Yang, Fei Wen, Erlei Wang

**Table of Contents**

## 1. The plane wave expansion method

In the linear limit Eq. (1) reduces to

$$i\frac{\partial U}{\partial Z} + \left(\frac{\partial^2}{\partial X^2} + \frac{\partial^2}{\partial Y^2}\right)U - VU = 0 \tag{S1}$$

The band structure of a periodic lattice can be found by substituting a solution of the form $U(X,Y,Z) = \Phi(X,Y)e^{-iqZ}$ into Eq. (S1). This results in the following eigenvalue equation

$$\left(\frac{\partial^2}{\partial X^2} + \frac{\partial^2}{\partial Y^2}\right)\Phi - V\Phi = -q\Phi \tag{S2}$$

Following Bloch's theorem, the eigenfunction $\Phi(r)$ in a periodic potential can be presented in the form of a product of a periodic function in space and a complex exponential: $\Phi(r)=\phi(r)\exp(ik\,r)$, where $\phi(r)$ is periodic, possessing the period of the lattice, and $k$ is its Bloch momentum. Invoking Fourier analysis, a periodic function can be expanded in terms of an infinite, discrete, sum of spatial harmonics: $\phi(r)=\Sigma h(G)\exp(iG\,r)$, where $G = b_1 P_1 + b_2 P_2$. $(P_1, P_2)$ are any integers and $(b_1, b_2)$ are the reciprocal basis vectors of the lattice. Thus, the eigenfunction $\Phi(r)$ can be written as $\Phi(r) = \sum_{G} h(G)e^{i(k+G)\cdot r}$. Similarly, the periodic potential can be expanded as

$$V(r) = \sum_{G} f(G)e^{iG\cdot r}, \text{ where } f(G) = \frac{1}{S_{cell}}\int_{cell} V(r)\exp(-iG\cdot r)ds \text{ is the expansion}$$

coefficient and $S_{cell} = (\sqrt{3}/2)$ is the area of the unit cell. Substituting the expressions of $\Phi(r)$ and $V(r)$ into Eq. (S2) leads to

$$\sum_{G'}\left[\delta(G-G')|k+G|^2 + f(G-G')\right]h(G') = qh(G) \tag{S3}$$

Typical examples of the linear lattice potential $V$ are depicted in Fig. S1. Given the potential, the eigenvalue equation (S3) can be solved numerically to obtain the lattice band structure. The results are shown in Fig. S2 for the potentials of Fig. S1. As can be seen, Dirac cones appear in these cases at the six corners of the Brillouin zone with eigenvalues, respectively, $q_D=115.644$, $q_D=-8.67$, $q_D=29.4445$, and $q_D=-41.81$.

## 2. Excitation of linear localized modes using numerical simulations

A circular defect at the origin is introduced by setting $V(X,Y)=V_d$, for $\sqrt{X^2+Y^2} \leq R$. Linear localized modes in these structures are discovered by using the finite difference beam propagation method (BPM) based on the evolution Eq. (S1), together with a transparent boundary condition. A source beam with phase varying as

$\exp(-iq_D Z)$: $S(X,Y,Z) = \exp\left[-\left(X^2+Y^2\right)\right]\exp(-iq_D Z)$, is actually launched in the defect waveguide. By changing the parameters ($R$ and $V_d$) of the defect, a situation arises in which the power in the waveguide monotonously grows with propagation distance, as shown in Fig. S3(a) (the initial stage) using the potential shown in Fig. S1(d), with $R=2$ and $V_d=-43.3$. This happens when synchronization of the waveguide eigenmode with the source beam is established, so that the eigenmode is always in-phase with the source and energy is extracted at every step of the propagation. This indicates the excitation of an eigenmode of the waveguide that has eigenvalue $q_D$ as the propagation constant. The field profile of this linear defect-guided mode is shown in Fig. S3(b). The BPM yields the space domain response $U(Z)$ directly. The spectral response $u(q)$ is subsequently obtained by the discretized Fourier transform from the spatial series $u(q) = \frac{1}{N}\sum_{n=0}^{N} U(n\Delta Z)e^{iqn\Delta Z}$, where $\Delta Z$ is step-size, $N$ is the number of steps, and $q$ (equivalent to spatial frequency) is the propagation constant. The spectrum of the eigenmode, obtained in this way, is shown in Fig. S3(c), which verifies that the propagation constant of the guided mode is truly centered at $q_D=-41.81$. In the second stage of evolution shown in Fig. S3(a), the source is switched off and amplitude of the eigenmode starts a propagation decline. In this free evolution stage, the electromagnetic power within the mode decays slowly and exponential according to the format as $P=P_0 e^{-\alpha Z}$. This law shows clearly that the instantaneous decay rate is $\alpha = -\frac{1}{P}\frac{dP}{dZ}$ and this can be calculated from slope of the power curve, after the evolution of beam power is numerically obtained. The instantaneous $\alpha$ value calculated in this way is shown in Fig. S3(d), as a function of the propagation distance.

The guided mode exists for a range of propagation constants around the Dirac point. Eigenmodes for propagation constants other than value for the Dirac point can be found in a similar way by adjusting the source beam to have other phase constants. The results are shown in Figs. S4(a)-S4(d), corresponding to the lattice potentials, respectively, of Figs. S1(a)-S1(d). They show parameter $V_d$ of the defect versus the propagation constant $q$ of the eigenmode for fixed $R$. However, as the propagation constant moves away from the Dirac point, the density of states increases in a linear fashion. Hence, there exists two loss mechanisms for the guided mode. If the propagation constant does not coincide with $q_D$ then, in addition to leakage $\alpha_c$ to the surrounding medium due to the finite lattice size (called losses associated with field penetration across boundary into the surrounding), there can be leakage $\alpha_s$ due to scattering into the continuum of states ( called losses associated with coupling into radiation modes). The net decay rate $\alpha=\alpha_c+\alpha_s$ increases with $|q-q_D|$. Leakage of the guided mode can be conveniently studied by the BPM. Suppose the beam power is $P_1$ at $Z_1$ and $P_2$ at $Z_2$, with the forms $P_1=P_0\exp(-\alpha Z_1)$, $P_2=P_0 \exp(-\alpha Z_2)$, then the average decay rate between $Z_1$ and $Z_2$ is $\alpha=\ln(P_1/P_2)/(Z_2-Z_1)$. The average loss rates $\alpha$, of the localized modes, extracted from the evolution of power are also shown in Fig. S4 as a function of the propagation constant $q$ using, respectively, the four different lattices. Minima of the loss rates $\alpha$ occur roughly around their corresponding Dirac propagation constants $q_D$, thus confirming the positions of the Dirac points.

## 3. The modified squared-operator iteration method for solitary waves

To find the Dirac-point solitons of Eq. (1), we seek a solution of the form $U(X,Y,Z) = \Phi(X,Y)e^{-iqZ}$. Given this substitution, Eq. (1) reduces to

$$\left(\frac{\partial^2}{\partial X^2} + \frac{\partial^2}{\partial Y^2}\right)\Phi - V_{NL}\Phi = -q\Phi \tag{S4}$$

where $V_{NL} = \dfrac{V_0}{V_0/V + |\Phi|^2}$ for the saturable nonlinearity and $V_{NL} = V - \sigma|\Phi|^2$ for a Kerr nonlinearity. Eq. (S4) is a static nonlinear Schrödinger equation that can be solved numerically by the modified squared-operator method for solitary wave solutions. The technique deployed here is to find solitary wave solutions of Eq. (S4) by iteration methods. For the fundamental soliton, it should be noted that $\Phi$ is real and $\partial\Phi/\partial X = \partial\Phi/\partial Y = 0$ at $(X, Y) = (0, 0)$, also $\Phi(\infty) = 0$. Progress towards a solution can then be made by supposing that an approximate real solution $\Phi_n$ exists, which is close to the exact solution $\Phi$. To obtain the next, iterative, form of the solution, $\Phi_{n+1}$, the following procedure is followed. First, express the exact solution as $\Phi = \Phi_n + \Delta\Phi$, where $\Delta\Phi \ll \Phi$ is the error term. Then substitute this expression into Eq. (S4) and expand it around $\Phi_n$. This leads to the linear inhomogeneous equation for the error $\Delta\Phi$: $L_0\Phi_n = -L_1\Delta\Phi$, which, in turn, gives $\Delta\Phi = -L_1^+ L_0\Phi_n$, where $L_1^+$ is the Hermitian of $L_1$. In the case of the fundamental soliton, $L_0 = \dfrac{\partial^2}{\partial x^2} + \dfrac{\partial^2}{\partial y^2} - V_{NL} + q$, $L_1 = L_0 - 2|\Phi|^2 \dfrac{\partial V_{NL}}{\partial |\Phi|^2}$, and the Hermitian of $L_1$ is $L_1$ itself. The approximate solution can then be updated to $\Phi_{n+1} = \Phi_n + \Delta\Phi$. This equation has a simpler appearance but converges very slowly. The situation can be improved by the introduction of the acceleration operator $M$, then $\Delta\Phi = -M^{-1}L_1^+ M^{-1}L_0\Phi_n \Delta t$, where $M = 30 - (\partial_{xx} + \partial_{yy})$ is a real-valued positive-definite Hermitian operator, and the step size $\Delta t$ controls the speed of convergence of the program. The convergence can be further speeded up by adopting the modified squared-operator iteration method. Using the forward Euler scheme, $\Phi_{n+1} = \Phi_n - [M^{-1}L_1^+ M^{-1}L_0\Phi - \alpha_n \langle G_n, L_1^+ M^{-1}L_0\Phi\rangle G_n]_{\Phi=\Phi_n} \Delta t$, where

$$\alpha_n = \frac{1}{\langle MG_n, G_n\rangle} - \frac{1}{\langle L_1 G_n, M^{-1}L_1 G_n\rangle \Delta t}, \quad G_n = \Phi_n - \Phi_{n-1}, \quad \langle F_1, F_2\rangle = \int_{-\infty}^{+\infty} F_1^+ \cdot F_2 \cdot dx.$$

The Dirac point can also support a vortex soliton of the form: $\Phi = f(r)e^{im\theta}$, where $f(r)=0$ at the beam center $r=0$ and $f(\infty)=0$ for a bright vortex soliton. The integer $m$ stands for a phase twist around the intensity ring and is usually called the *winding number*. Typical Dirac-point solitons (fundamental and vortex) and the power $P = \iint |\Phi|^2 \, dXdY$ conveyed by the solitons versus the eigenvalue $q$ are shown in Figs. S5-S8 respectively for the focusing/defocusing saturable and Kerr nonlinearities. The gray scale in the background of Figs. S5(c, d) - S8(c, d) indicates the level of linear losses $\alpha$ of the wave (as given in Fig. S4) in the lattice potential, which marks roughly the range of propagation constant within which the associated Dirac-point solitons can exist. The Dirac-point solitons are found to exist within the ranges shaded light gray, where the level of the linear decay rate $\alpha$ is low. The dotted vertical lines indicate the positions of the Dirac points.

## 4. The linear stability analysis

To study the stability of these Dirac-point solitons, a perturbation of the form $U = e^{-iqZ}\left[\Phi + (v-w)e^{\lambda Z} + (v+w)^* e^{\lambda^* Z}\right]$, is invoked, where $v, w \ll 1$. For Kerr nonlinearity substituting this perturbation into

$$i\frac{\partial U}{\partial Z} + \left(\frac{\partial^2}{\partial X^2} + \frac{\partial^2}{\partial Y^2}\right)U - VU + \sigma |U|^2 U = 0$$ and then linearizing, results in

$$q(v-w)e^{\lambda Z} + q(v+w)^* e^{\lambda^* Z} + i(v-w)e^{\lambda Z}\lambda + i(v+w)^* e^{\lambda^* Z}\lambda^* + \left(\frac{\partial^2}{\partial X^2} + \frac{\partial^2}{\partial Y^2}\right)(v-w)e^{\lambda Z}$$

$$+ \left(\frac{\partial^2}{\partial X^2} + \frac{\partial^2}{\partial Y^2}\right)(v+w)^* e^{\lambda^* Z} - V(v-w)e^{\lambda Z} - V(v+w)^* e^{\lambda^* Z} + 2\sigma \Phi^* \Phi (v-w)e^{\lambda Z}$$

$$+ 2\sigma \Phi^* \Phi (v+w)^* e^{\lambda^* Z} + \sigma \Phi^2 (v^* - w^*)e^{\lambda^* Z} + \sigma \Phi^2 (v+w)e^{\lambda Z} = 0$$

Separate the two groups with respectively factors of $e^{\lambda Z}$ and $e^{\lambda^* Z}$

$$q(v-w) + i(v-w)\lambda + \left(\frac{\partial^2}{\partial X^2} + \frac{\partial^2}{\partial Y^2}\right)(v-w) - V(v-w) + 2\sigma \Phi^* \Phi (v-w) + \sigma \Phi^2 (v+w) = 0$$

$$q(v+w)^* + i(v+w)^* \lambda^* + \left(\frac{\partial^2}{\partial X^2} + \frac{\partial^2}{\partial Y^2}\right)(v+w)^* - V(v+w)^* + 2\sigma \Phi^* \Phi (v+w)^* + \sigma \Phi^2 (v^* - w^*) = 0$$

Taking complex conjugate of the second equation

$$q(v-w) + i(v-w)\lambda + \Delta(v-w) - V(v-w) + 2\sigma \Phi^* \Phi (v-w) + \sigma \Phi^2 (v+w) = 0$$

$$q(v+w) - i(v+w)\lambda + \Delta(v+w) - V(v+w) + 2\sigma \Phi^* \Phi (v+w) + \sigma \Phi^{*2} (v-w) = 0$$

where the operator is $\Delta = \frac{\partial^2}{\partial X^2} + \frac{\partial^2}{\partial Y^2}$. The above equation can be rearranged into

$$\tfrac{1}{2}\sigma\left(\Phi^{*2}-\Phi^2\right)v+\left(\Delta+q-V+2\sigma|\Phi|^2-\tfrac{1}{2}\sigma\Phi^2-\tfrac{1}{2}\sigma\Phi^{*2}\right)w=i\lambda v$$

$$\left(\Delta+q-V+2\sigma|\Phi|^2+\tfrac{1}{2}\sigma\Phi^2+\tfrac{1}{2}\sigma\Phi^{*2}\right)v+\tfrac{1}{2}\sigma\left(\Phi^2-\Phi^{*2}\right)w=i\lambda w$$

which is a linear eigenvalue problem $L\Psi=\lambda\Psi$ with $\Psi$ being the transpose of $(v, w)$ and

$$L=-i\begin{pmatrix}\tfrac{1}{2}\sigma\left(\Phi^{*2}-\Phi^2\right) & \Delta+q-V+2\sigma|\Phi|^2-\tfrac{1}{2}\sigma\left(\Phi^2+\Phi^{*2}\right)\\ \Delta+q-V+2\sigma|\Phi|^2+\tfrac{1}{2}\sigma\left(\Phi^2+\Phi^{*2}\right) & \tfrac{1}{2}\sigma\left(\Phi^2-\Phi^{*2}\right)\end{pmatrix}$$

For saturable nonlinearity substituting the perturbation

$$U(X,Y,Z)=e^{-iqZ}\left[\Phi+(v-w)e^{\lambda Z}+(v+w)^* e^{\lambda^* Z}\right]\text{ into}$$

$$i\frac{\partial U}{\partial Z}+\left(\frac{\partial^2}{\partial X^2}+\frac{\partial^2}{\partial Y^2}\right)U-\frac{V_0}{V_0/V+|U|^2}U=0\text{ and then linearizing, results in}$$

$$q(v-w)e^{\lambda Z}+q(v+w)^* e^{\lambda^* Z}+i(v-w)\lambda e^{\lambda Z}+i(v+w)^*\lambda^* e^{\lambda^* Z}+\left(\frac{\partial^2}{\partial X^2}+\frac{\partial^2}{\partial Y^2}\right)(v-w)e^{\lambda Z}$$

$$+\left(\frac{\partial^2}{\partial X^2}+\frac{\partial^2}{\partial Y^2}\right)(v+w)^* e^{\lambda^* Z}-\frac{V_0}{V_0/V+|\Phi|^2}\left[(v-w)e^{\lambda Z}+(v+w)^* e^{\lambda^* Z}\right]$$

$$+\frac{V_0\Phi}{\left(V_0/V+|\Phi|^2\right)^2}\left[\Phi(v-w)^* e^{\lambda^* Z}+\Phi(v+w)e^{\lambda Z}+\Phi^*(v-w)e^{\lambda Z}+\Phi^*(v+w)^* e^{\lambda^* Z}\right]=0$$

Separate the two groups with respectively factors of $e^{\lambda Z}$ and $e^{\lambda^* Z}$ and take complex conjugate of the second equation

$$q(v-w)+i(v-w)\lambda+\left(\frac{\partial^2}{\partial X^2}+\frac{\partial^2}{\partial Y^2}\right)(v-w)-\frac{V_0}{\left(V_0/V+|\Phi|^2\right)^2}\left((v-w)V_0/V-\Phi^2(v+w)\right)=0$$

$$q(v+w)-i(v+w)\lambda+\left(\frac{\partial^2}{\partial X^2}+\frac{\partial^2}{\partial Y^2}\right)(v+w)-\frac{V_0}{\left(V_0/V+|\Phi|^2\right)^2}\left((v+w)V_0/V-\Phi^{*2}(v-w)\right)=0$$

Rearranging into

$$\tfrac{1}{2}\frac{V_0}{\left(V_0/V+|\Phi|^2\right)^2}\left(\Phi^{*2}-\Phi^2\right)v+qw+\Delta w-wV_0/V\frac{V_0}{\left(V_0/V+|\Phi|^2\right)^2}-\tfrac{1}{2}\frac{V_0}{\left(V_0/V+|\Phi|^2\right)^2}\left(\Phi^2+\Phi^{*2}\right)w=iv\lambda$$

$$qv+\Delta v-vV_0/V\frac{V_0}{\left(V_0/V+|\Phi|^2\right)^2}+\tfrac{1}{2}\frac{V_0}{\left(V_0/V+|\Phi|^2\right)^2}\left(\Phi^2+\Phi^{*2}\right)v+\tfrac{1}{2}\frac{V_0}{\left(V_0/V+|\Phi|^2\right)^2}\left(\Phi^2-\Phi^{*2}\right)w=iw\lambda$$

which is a linear eigenvalue problem $L\Psi=\lambda\Psi$ with $\Psi$ being the transpose of $(v, w)$ and

$$L=-i\begin{pmatrix}\tfrac{1}{2}\frac{V_0}{\left(V_0/V+|\Phi|^2\right)^2}\left(\Phi^{*2}-\Phi^2\right) & q+\Delta-V_0/V\frac{V_0}{\left(V_0/V+|\Phi|^2\right)^2}-\tfrac{1}{2}\frac{V_0}{\left(V_0/V+|\Phi|^2\right)^2}\left(\Phi^2+\Phi^{*2}\right)\\ q+\Delta-V_0/V\frac{V_0}{\left(V_0/V+|\Phi|^2\right)^2}+\tfrac{1}{2}\frac{V_0}{\left(V_0/V+|\Phi|^2\right)^2}\left(\Phi^2+\Phi^{*2}\right) & \tfrac{1}{2}\frac{V_0}{\left(V_0/V+|\Phi|^2\right)^2}\left(\Phi^2-\Phi^{*2}\right)\end{pmatrix}$$

The linear stability eigenvalue problem $L\Psi=\lambda\Psi$ is solved by the numerical iteration method. The real part of the perturbation growth rate Re($\lambda$) versus the propagation constant $q$ of the soliton is plotted in Figs. S5(d)-S8(d) respectively for the four different lattices shown in Figs. S1(a)-S1(d). In a saturable, self-focusing, lattice (Fig. S5), growth rates exceed 10, both the fundamental soliton and the first vortex soliton are unstable. In a saturable self-defocusing lattice (Fig. S6), growth rates are of the order of 1, both the fundamental soliton and the first vortex soliton are weakly unstable. In Kerr lattices (Figs. S7 and S8), the fundamental soliton is unstable for self-focusing nonlinearity and stable for self-defocusing nonlinearity. The stability of Dirac-point soliton can be checked numerically using the split-step Fourier method. Simulated evolution scenarios of typical stable and unstable solitons are shown in Figs. S9 and S10 respectively, confirming the stability analysis. The propagation distance of the stable soliton (Fig. S10) is about three orders of magnitude larger than that of the unstable soliton (Fig. S9).

Propagation distance of the Dirac-point soliton is not infinite when loss is in existence, even for the stable case. Owing to the slowly decaying tail of the Dirac-point soliton itself, losses associated with field penetration across boundary into the surrounding is sensitive to the lattice size. As the Dirac-point soliton breaks down, or as it loses power to the surrounding medium, its amplitude reduces. This alters parameters of the nonlinearity-induced defect, and, in turn, shifts its propagation constant. This is a process of self-propagation-constant shift that the Dirac-point soliton undergoes in propagation, an analogue of self-frequency shift of a temporal soliton. As the propagation constant of the soliton deviates from the value of the Dirac point, losses associated with coupling into radiation modes arise and accelerate degradation of the soliton. Therefore, even for a stable soliton, the losses will sooner, or later, breakup the balance between contraction and diffraction, and eventually diminish the soliton. As such, the propagation distance of a stable Dirac-point soliton is restricted by the finite lattice size. On the other hand, a gap soliton is hardly influenced by a distanced boundary because its tail decays exponentially in space. At the Dirac point the density of radiation states vanishes, any residual values of loss rate at this point are entirely due to the finite lattice size, which can be made as small as desired by increasing the boundary surrounding the soliton. Therefore, the propagation distance of the stable Dirac-point soliton can be extended to almost as long as desired.

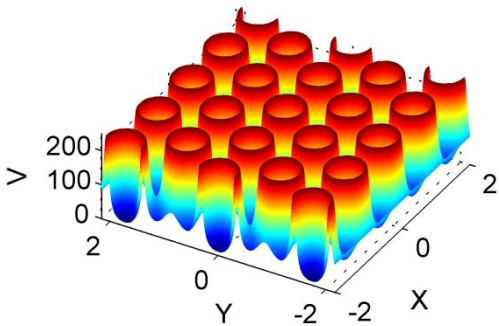 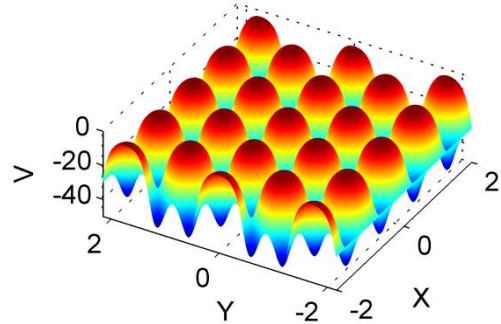

(a)                                            (b)

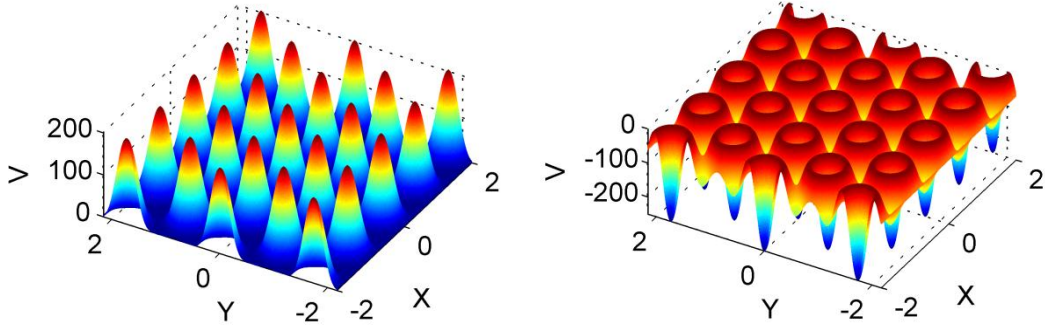

(c)                                                  (d)

Fig. S1. Four different lattice potentials. (a, b) Index potential of the photorefractive material (saturable nonlinearity) $V = \dfrac{V_0}{1+I_0\left[\chi+\cos(b_1\cdot r)+\cos(b_2\cdot r)+\cos(b_3\cdot r)\right]^2}$ for $I_0=2$, $\chi=0$, $V_0=250$ (a, self-focusing) and $I_0=1$, $\chi=3$, $V_0=-150$ (b, self-defocusing). (c, d) Index potential $V = V_0\left[\chi+\cos(b_1\cdot r)+\cos(b_2\cdot r)+\cos(b_3\cdot r)\right]^2$ of the Kerr nonlinear medium for $V_0=10$, $\chi=3/2$ (c) and $V_0=-35$, $\chi=-1/3$ (d). In the cases of (b) and (c) the potentials exhibit absolute index maxima on lattice sites, while in the cases of (a) and (d) the potentials exhibit absolute index minima on lattice sites.

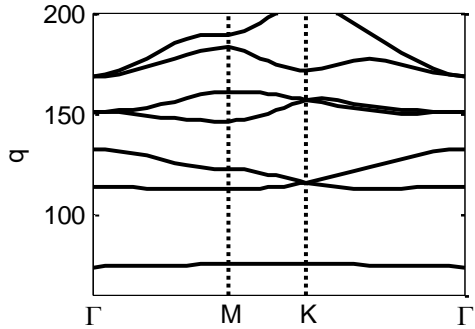 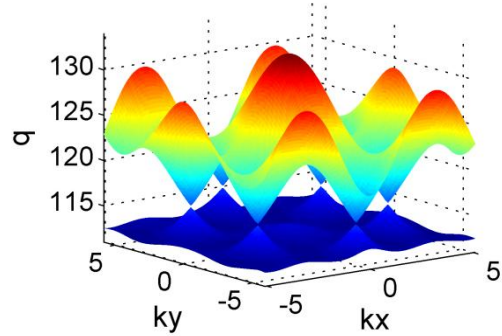

(a)

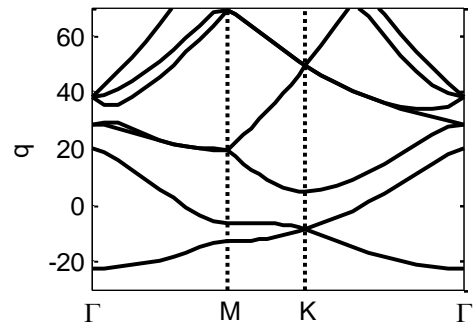 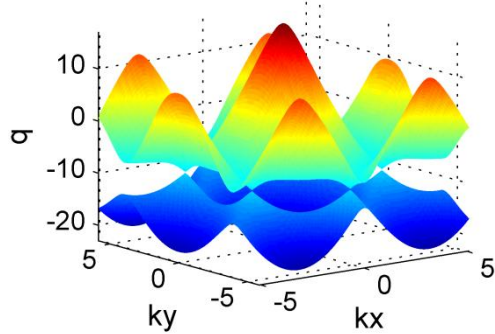

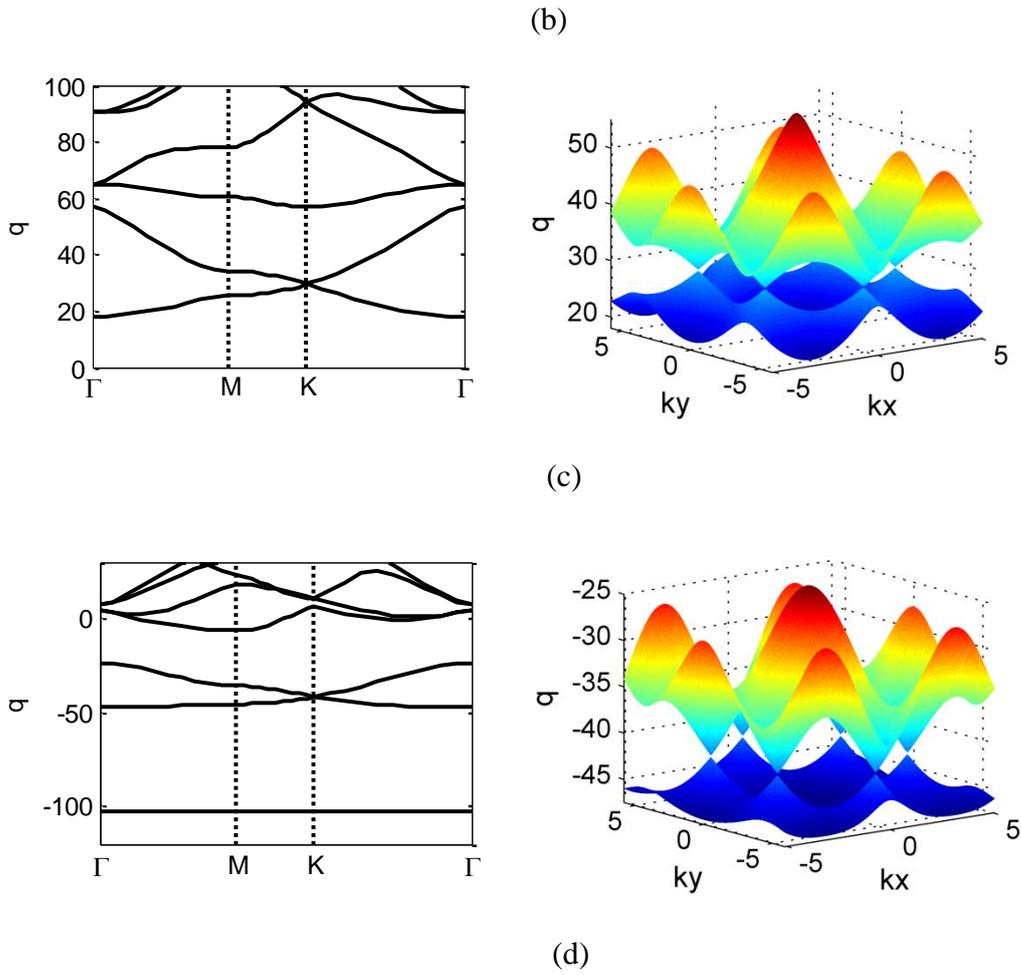

Fig. S2. Band structures (left) and the enlarged 3D views of linear Dirac cones around the Dirac point (right) of four different optical lattices. (a-d) correspond to the lattices shown in Figs. S1(a)-S1(d): (a) $I_0=2$, $\chi=0$, $V_0=250$, (b) $I_0=1$, $\chi=3$, $V_0=-150$, (c) $V_0=10$, $\chi=3/2$, and (d) $V_0=-35$, $\chi=-1/3$.

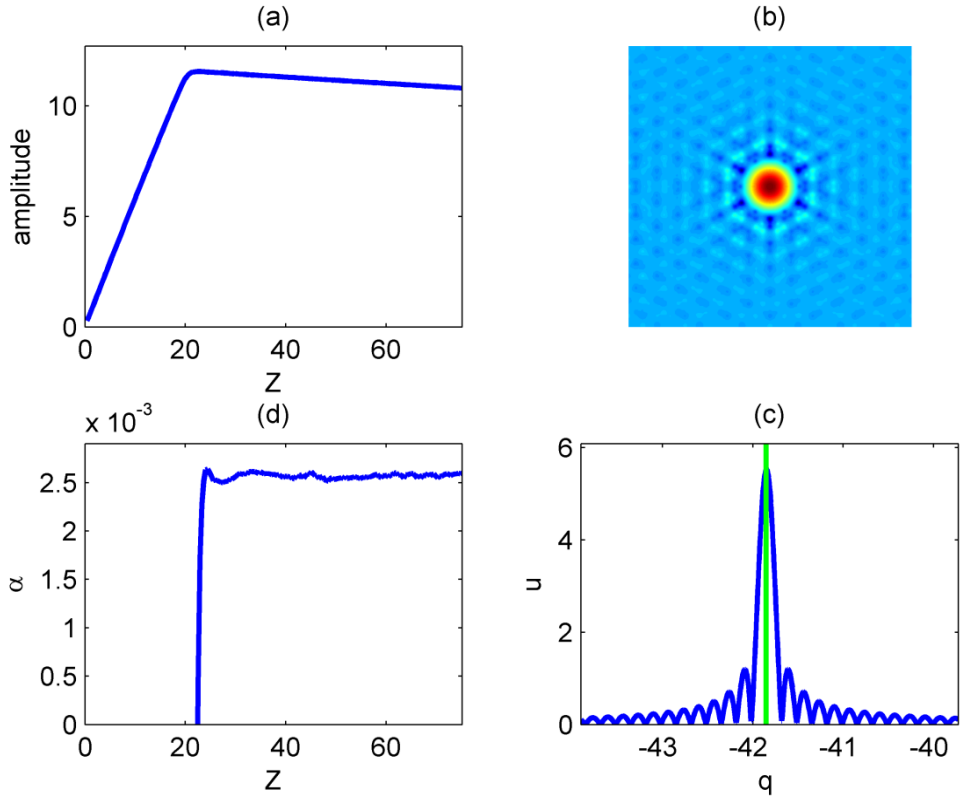

Fig. S3. Mode in a waveguide formed by a defect $V(X,Y)=V_d$ for $\sqrt{X^2+Y^2} \leq R$ in the potential of Fig. S1(d). Parameters of the defect are $R=2$ and $V_d=-43.3$. (a) Evolution of amplitude of the excited eigenmode. Between $Z=0$ and 20 a source beam with a phase constant $q_D=-41.81$ is on, i.e., the field evolves according to

$$i\frac{\partial U}{\partial Z}+\left(\frac{\partial^2}{\partial X^2}+\frac{\partial^2}{\partial Y^2}\right)U-VU=iS \text{ with } S=e^{-(X^2+Y^2)}e^{-iq_D Z}.$$

In the second stage ($Z>20$) the source beam is turned off and the eigenmode evolves freely on its own in the waveguide. (b) Profile of the eigenmode of the waveguide excited by the source beam. (c) The propagation constant spectrum of the excited mode. The green vertical line indicates the position of the Dirac point. (d) The instantaneous decay rate $\alpha$ of the excited mode as it propagates down the waveguide.

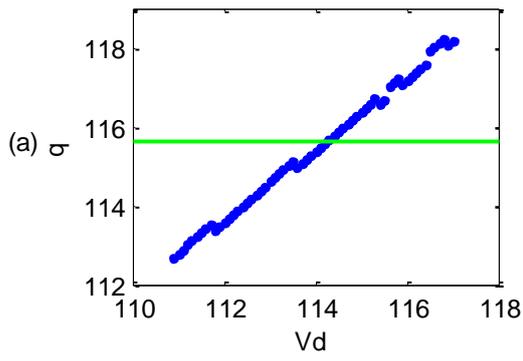
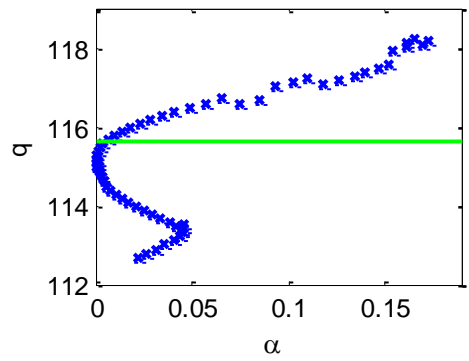

(a)

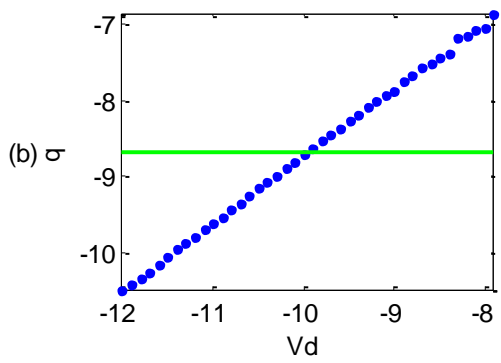
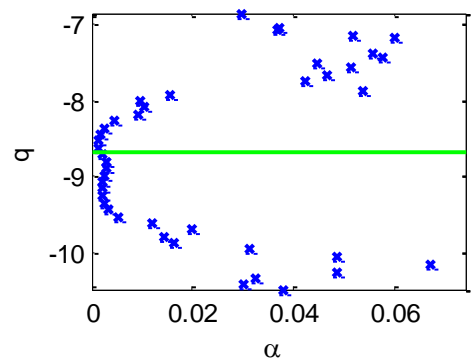

(b)

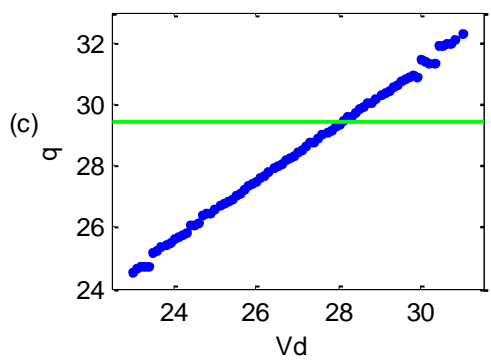
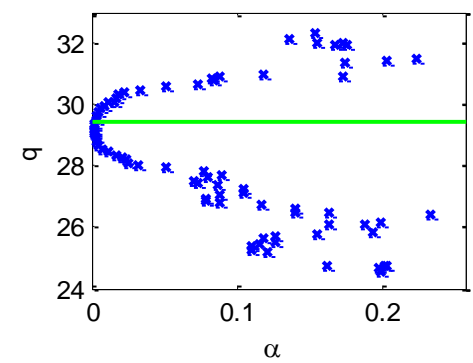

(c)

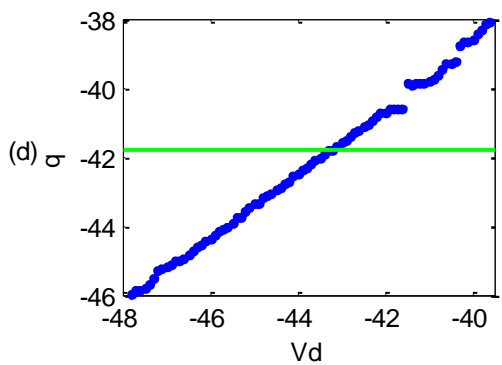
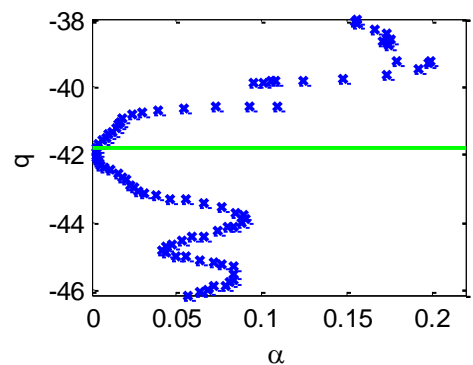

(d)

Fig. S4. (a-d) Parameter $V_d$ of the defect (left) and the average power loss rate $\alpha$ of the modes (right) versus the eigen propagation constant $q$ of the localized mode for respectively the lattices shown in Figs. S1(a)-S1(d). The computational domain is taken as a square of $-15 < X, Y < 15$ (a, b) or $-10 < X, Y < 10$ (c, d), discretized by 601 points along each dimension. The green horizontal lines indicate the positions of the Dirac points.

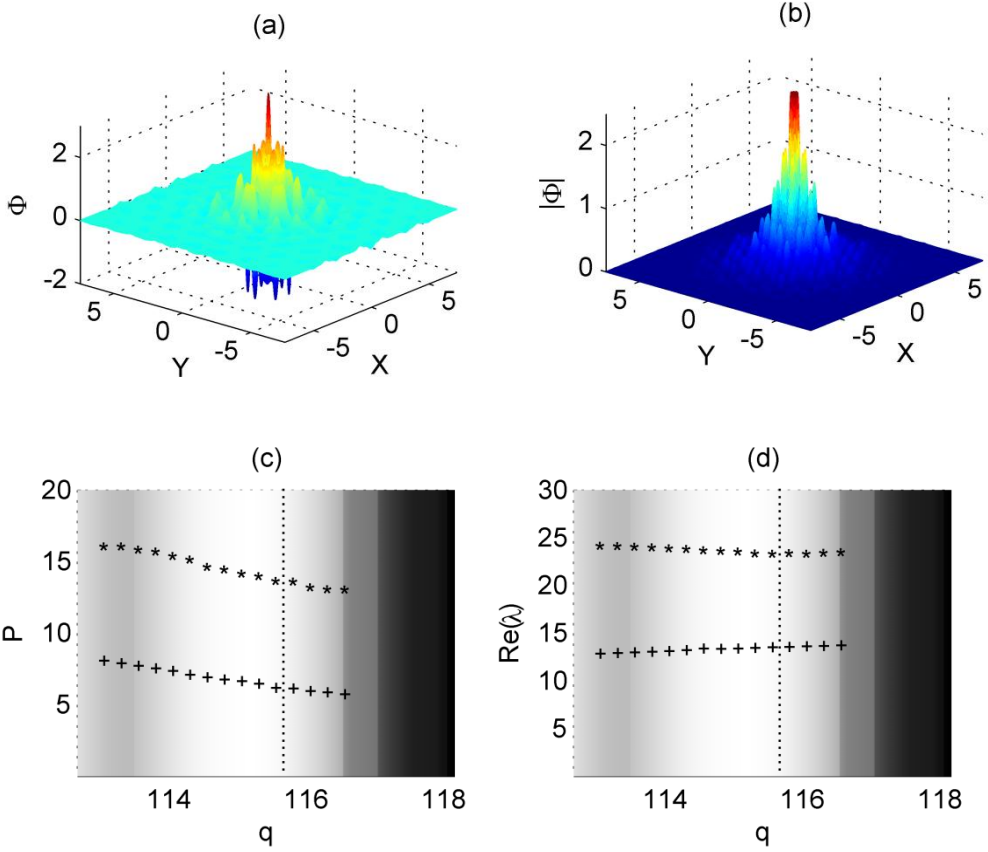

Fig. S5. Dirac-point solitons in a saturable self-focusing lattice. The lattice potential is shown in Fig. S1(a): $V = \dfrac{V_0}{1+I_0\left[\chi+\cos(b_1\cdot r)+\cos(b_2\cdot r)+\cos(b_3\cdot r)\right]^2}$ for $I_0=2$, $\chi=0$, $V_0=250$. (a, b) The field profiles of the fundamental soliton (a) and the first vortex soliton ($m=1$) (b) at the Dirac point $q_D=115.644$. The computational domain is taken as a square of $-7.5 < X, Y < 7.5$, discretized by 512 points along each dimension. The initial condition is taken as $\Phi=550\,\mathrm{sech}\left(\dfrac{20}{3}\sqrt{X^2+Y^2}\right)\cos\left(\dfrac{20}{3}\sqrt{X^2+Y^2}\right)(X^2+Y^2)$ and $\Phi=100\,\mathrm{sech}\left(4\sqrt{X^2+Y^2}\right)(X+iY)$ respectively. (c, d) Power $P$ (c) and the real part of the perturbation growth rate $\mathrm{Re}(\lambda)$ (d) versus the propagation constant $q$ of the soliton, where "*" represents the fundamental soliton and "+" represents the vortex

soliton. The dotted vertical line indicates the position of the Dirac point.

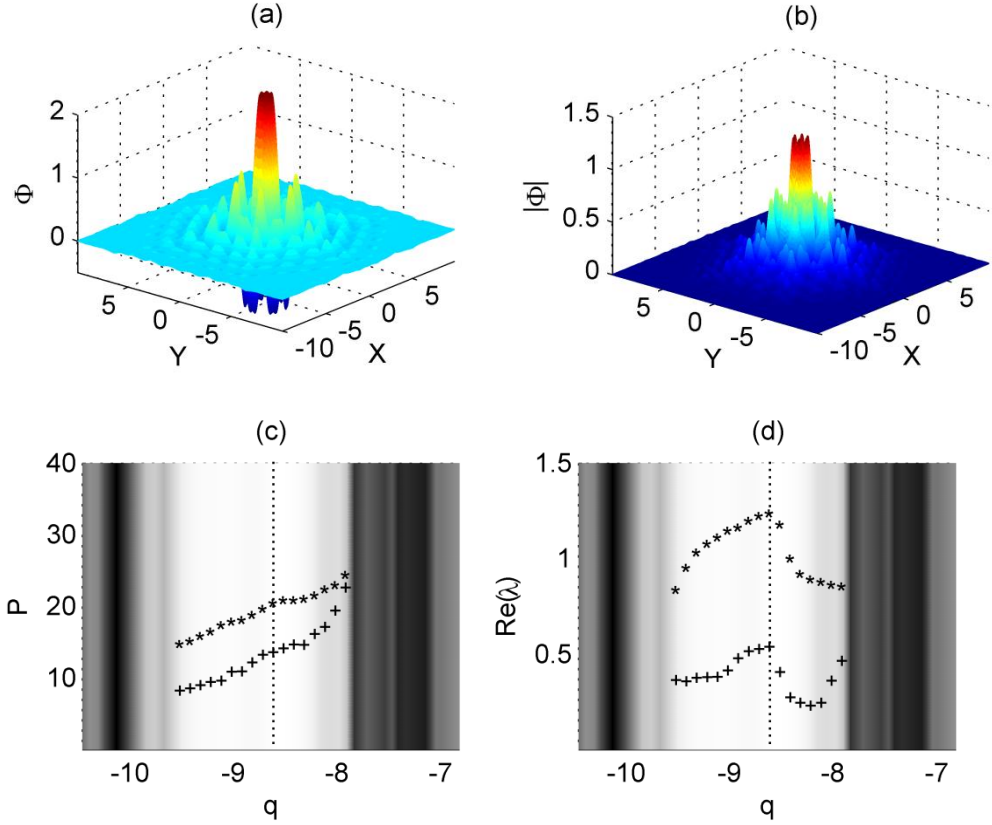

Fig. S6. Dirac-point solitons in a saturable self-defocusing lattice. The lattice potential is shown in Fig. S1(b): $V = \dfrac{V_0}{1+I_0[\chi+\cos(b_1\cdot r)+\cos(b_2\cdot r)+\cos(b_3\cdot r)]^2}$ for $I_0=1$, $\chi=3$, $V_0=-150$. (a, b) The field profiles of the fundamental soliton (a) and the first vortex soliton ($m=1$) (b) at the Dirac point $q_D=-8.67$. The computational domain is taken as a square of $-10 < X, Y < 10$, discretized by 512 points along each dimension. The initial condition is taken respectively as $\Phi=50\text{sech}\left(4\sqrt{X^2+Y^2}\right)$ and $\Phi=400\text{sech}\left(8\sqrt{X^2+Y^2}\right)(X+iY)$. (c, d) Power $P$ (c) and the real part of the perturbation growth rate $\text{Re}(\lambda)$ (d) versus the propagation constant $q$ of the soliton, where "*" represents the fundamental soliton and "+" represents the vortex soliton. The dotted vertical line indicates the position of the Dirac point.

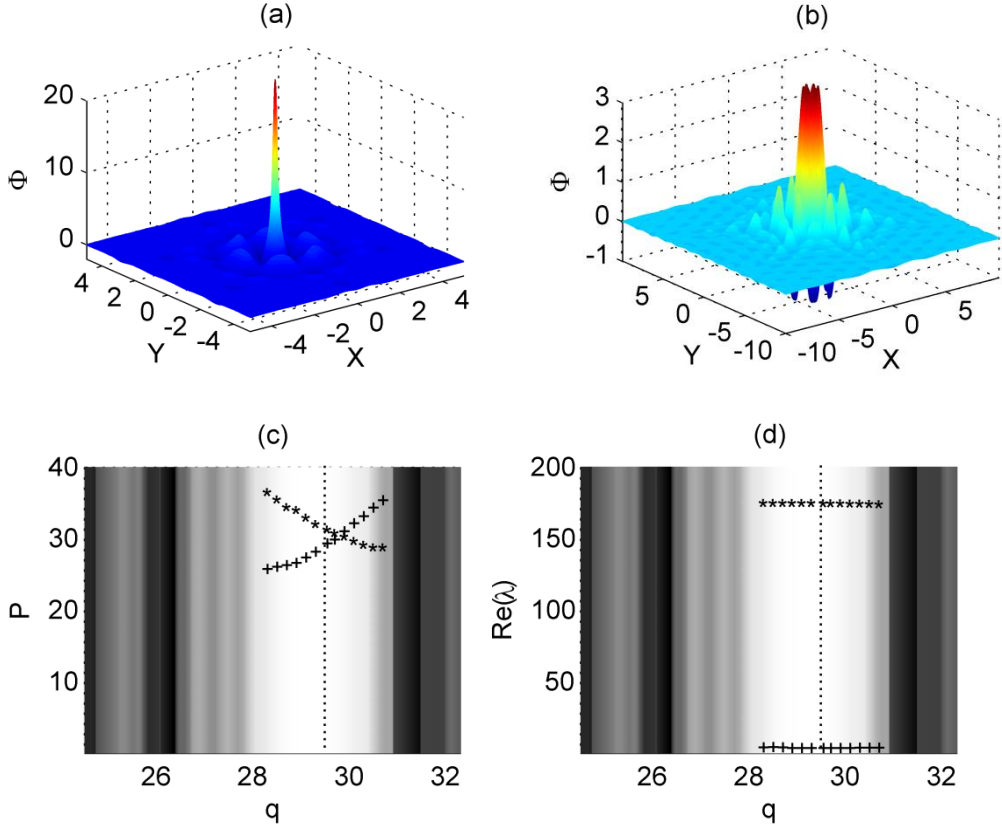

Fig. S7. Dirac-point solitons in Kerr nonlinear media. The lattice potential is shown in Fig. S1(c): $V = V_0[\chi + \cos(b_1 \cdot r) + \cos(b_2 \cdot r) + \cos(b_3 \cdot r)]^2$ for $V_0$=10, $\chi$=3/2, which exhibits absolute index maxima on lattice sites. (a, b) The field profiles of the fundamental solitons in a Kerr self-focusing ($\sigma$=1) (a) and self-defocusing ($\sigma$=−1) (b) lattice at the Dirac point $q_D$=29.445. The computational domain is taken as a square of respectively $-5 < X, Y < 5$ and $-10 < X, Y < 10$, discretized by 512 points along each dimension. The initial condition is respectively taken as $\Phi = 12\,\text{sech}\left(4\sqrt{X^2+Y^2}\right)$ for the self-focusing case and $\Phi = -30\sqrt{X^2+Y^2}\,\text{sech}\left(4\sqrt{X^2+Y^2}\right)\cos\left(5\sqrt{X^2+Y^2}\right)$ for the self-defocusing case. (c, d) Power $P$ (c) and real part of the perturbation growth rate Re($\lambda$) (d) versus the propagation constant $q$ of the soliton, where "*" corresponds to the self-focusing case and "+" corresponds to the self-defocusing case. The dotted vertical line indicates the position of the Dirac point.

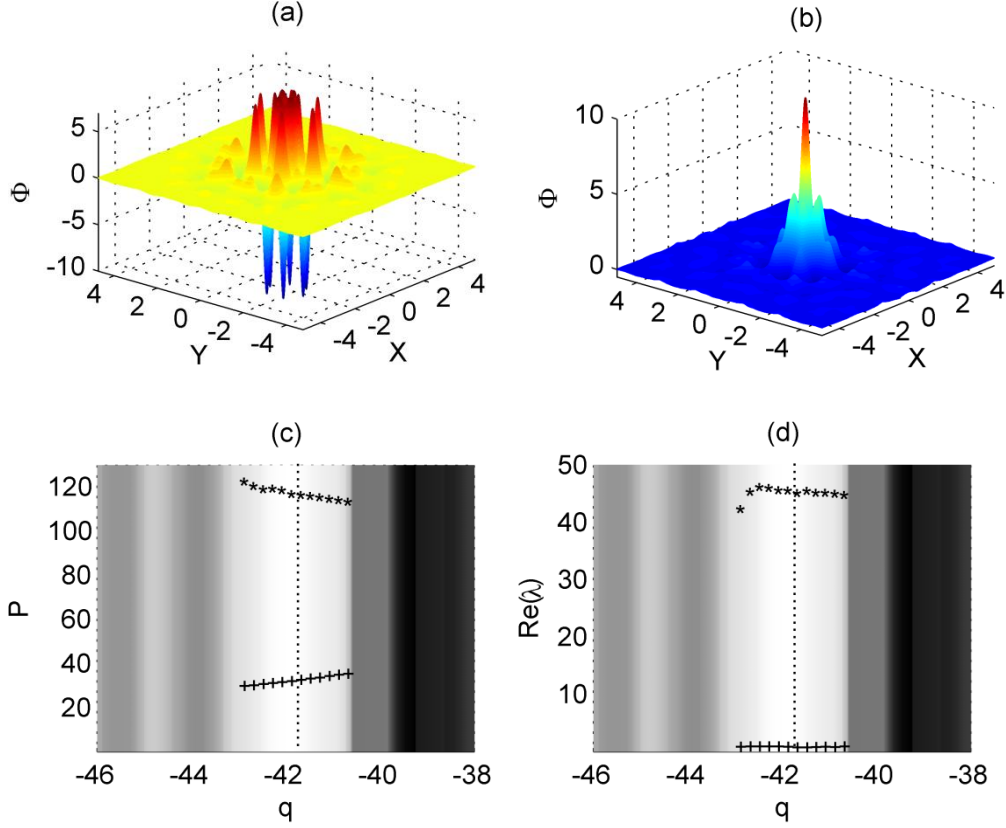

Fig. S8. Dirac-point solitons in Kerr nonlinear media. The lattice potential is shown in Fig. S1(d): $V = V_0[\chi + \cos(b_1 \cdot r) + \cos(b_2 \cdot r) + \cos(b_3 \cdot r)]^2$ for $V_0 = -35$, $\chi = -1/3$, which exhibits absolute index minima on lattice sites. (a, b) The field profiles of the fundamental solitons in a Kerr self-focusing ($\sigma = 1$) (a) and self-defocusing ($\sigma = -1$) (b) lattice at the Dirac point $q_D = -41.81$. The computational domain is taken as a square of $-5 < X, Y < 5$, discretized by 1024 points along each dimension. The initial condition is respectively $\Phi = -10(0.1 + 5\sqrt{X^2 + Y^2})\text{sech}(4\sqrt{X^2 + Y^2})\cos(8\sqrt{X^2 + Y^2})$ and $\Phi = 10\text{sech}(4\sqrt{X^2 + Y^2})$ for the two Dirac-point solitons. (c, d) Power $P$ (c) and real part of the perturbation growth rate $\text{Re}(\lambda)$ (d) versus the propagation constant $q$ of the soliton, where "*" corresponds to the self-focusing case and "+" corresponds to the self-defocusing case. The dotted vertical line indicates the position of the Dirac point.

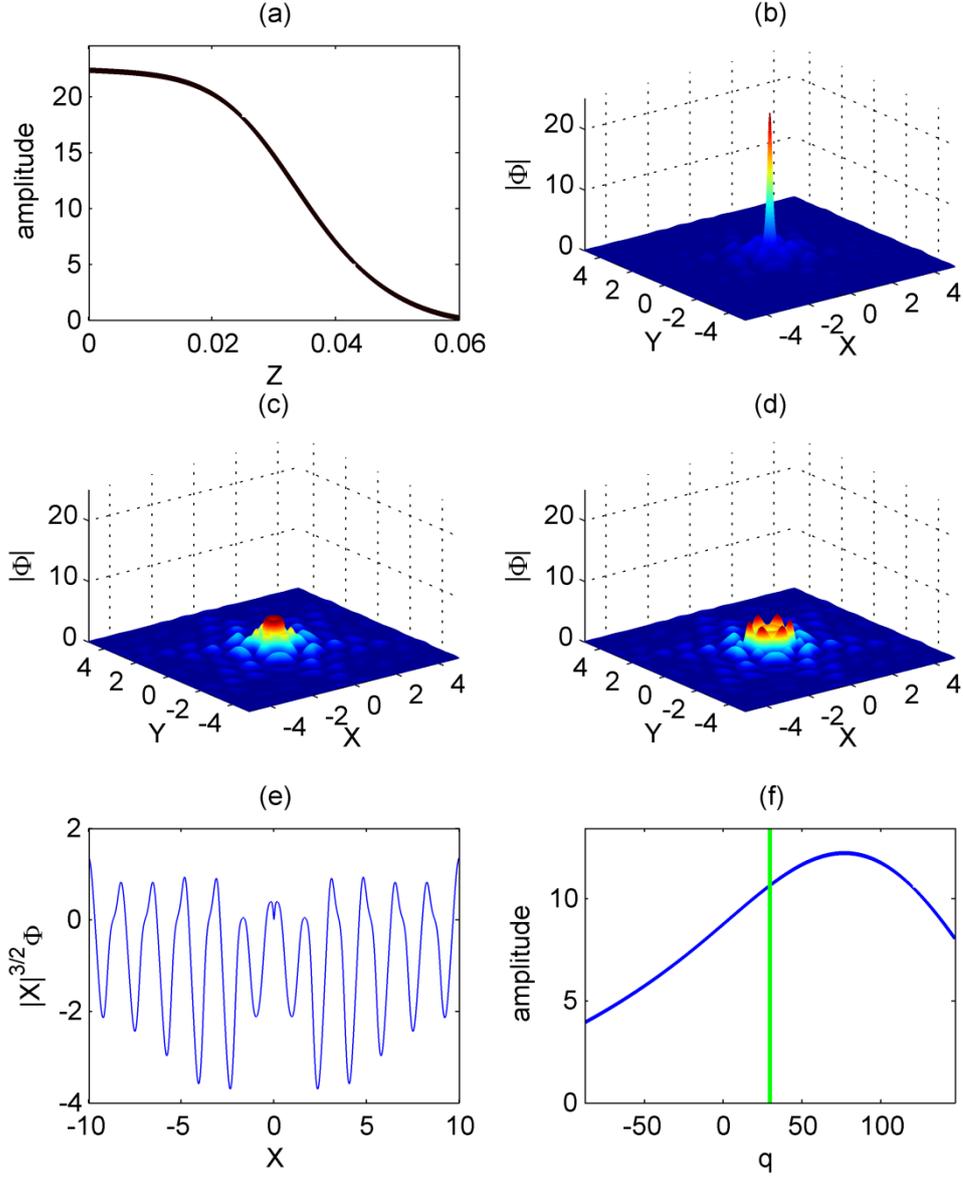

Fig. S9. Breakdown of the Dirac-point soliton in a Kerr self-focusing lattice. The lattice potential is shown in Fig. S1(c): $V = V_0 \left[ \chi + \cos(b_1 \cdot r) + \cos(b_2 \cdot r) + \cos(b_3 \cdot r) \right]^2$ for $V_0=10$, $\chi=3/2$. The initial profile of the soliton is shown in Fig. S7(a). The computational domain is taken as a square of $-10 < X, Y < 10$, discretized by 1024 points along each dimension. (a) Evolution of amplitude $|\Phi(0,0)|$ of the fundamental soliton in propagation. (b-d) The $|\Phi|$ field (zoomed in to $-5 < X, Y < 5$) of the soliton at respectively $Z=0.006, 0.048, 0.06$. (e) Product of the initial $\Phi$ and $r^{3/2}$ on the $X$ axis. (f) Propagation constant spectrum of the soliton. The green vertical line indicates the position of the Dirac point.

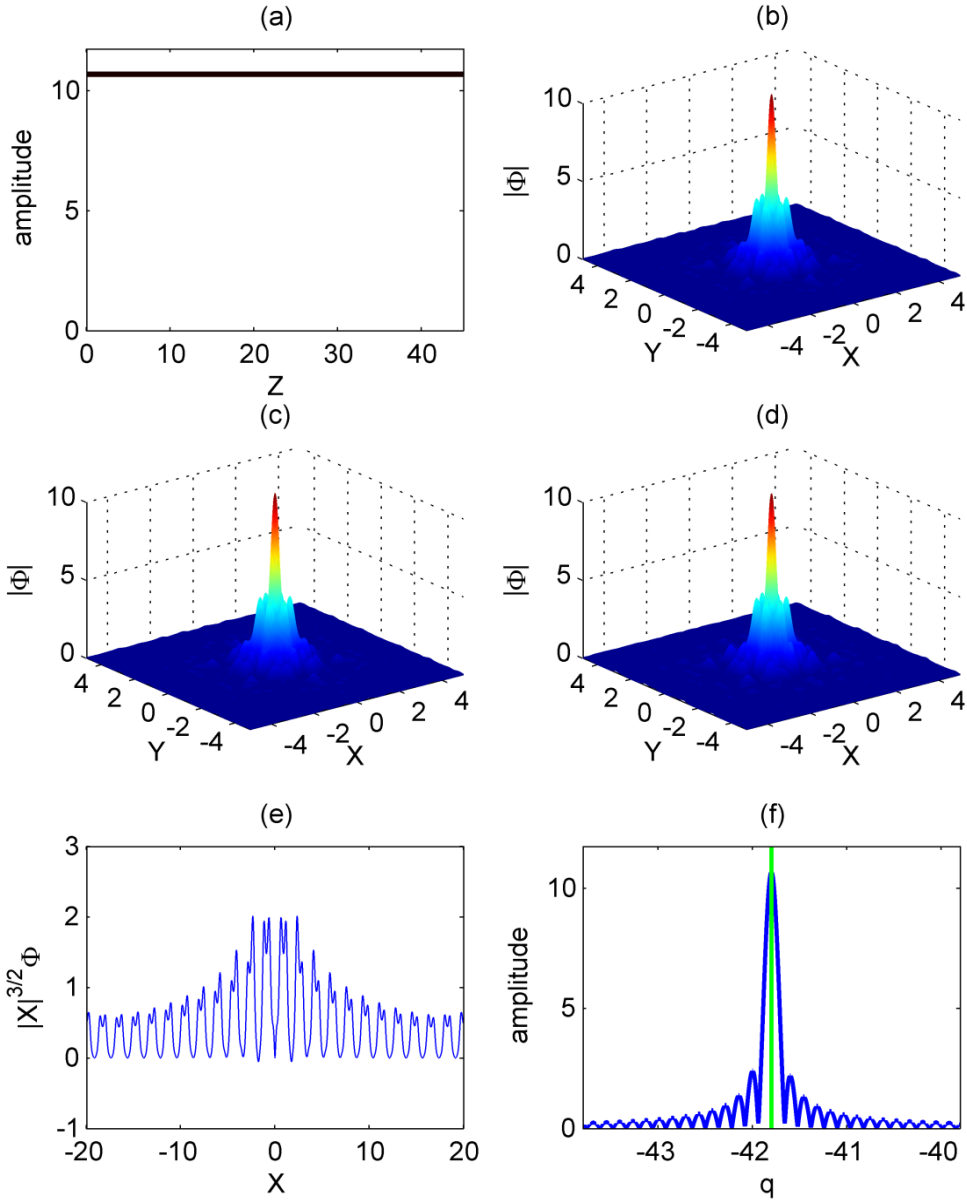

Fig. S10. Dynamics of the Dirac-point soliton in a Kerr self-defocusing lattice. The lattice potential is shown in Fig. S1(d): $V = V_0[\chi + \cos(b_1 \cdot r) + \cos(b_2 \cdot r) + \cos(b_3 \cdot r)]^2$ for $V_0=-35$, $\chi=-1/3$. The initial profile of the soliton is shown in Fig. S8(b). The computational domain is taken as a square of $-15 < X, Y < 15$, discretized by 1024 points along each dimension. (a) Evolution of amplitude $|\Phi(0,0)|$ of the fundamental soliton in propagation. (b-d) The $|\Phi|$ field (zoomed in to $-5 < X, Y < 5$) of the soliton at respectively $Z=4.5, 22.5, 45$. (e) Product of the initial $\Phi$ and $r^{3/2}$ on the X axis. (f) Propagation constant spectrum of the soliton. The green vertical line indicates the position of the Dirac point.